\pdfoutput=1

\documentclass[preprint,aps,pra]{revtex4}

\usepackage{times}
\usepackage{graphicx}
\usepackage{epsfig}
\usepackage{amsfonts}
\usepackage{amsmath}
\usepackage{amssymb}
\usepackage{color}
\usepackage{multirow}
\usepackage[colorlinks=true,citecolor=blue,urlcolor=blue]{hyperref}

\begin{document}


\title{The unspeakable why}


\author{Ad\'an~Cabello}
 \affiliation{Departamento de F\'{\i}sica Aplicada II, Universidad de
 Sevilla, E-41012 Sevilla, Spain}


\date{\today}


\begin{abstract}
 For years, the biggest unspeakable in quantum theory has been why quantum theory and what is quantum theory telling us about the world. Recent efforts are unveiling a surprisingly simple answer. Here we show that two characteristic limits of quantum theory, the maximum violations of Clauser-Horne-Shimony-Holt and Klyachko-Can-Binicio\u{g}lu-Shumovsky inequalities, are enforced by a simple principle. The effectiveness of this principle suggests that non-realism is the key that explains why quantum theory.
\end{abstract}


\maketitle


\begin{figure}[t]
\centering
\includegraphics[scale=0.70]{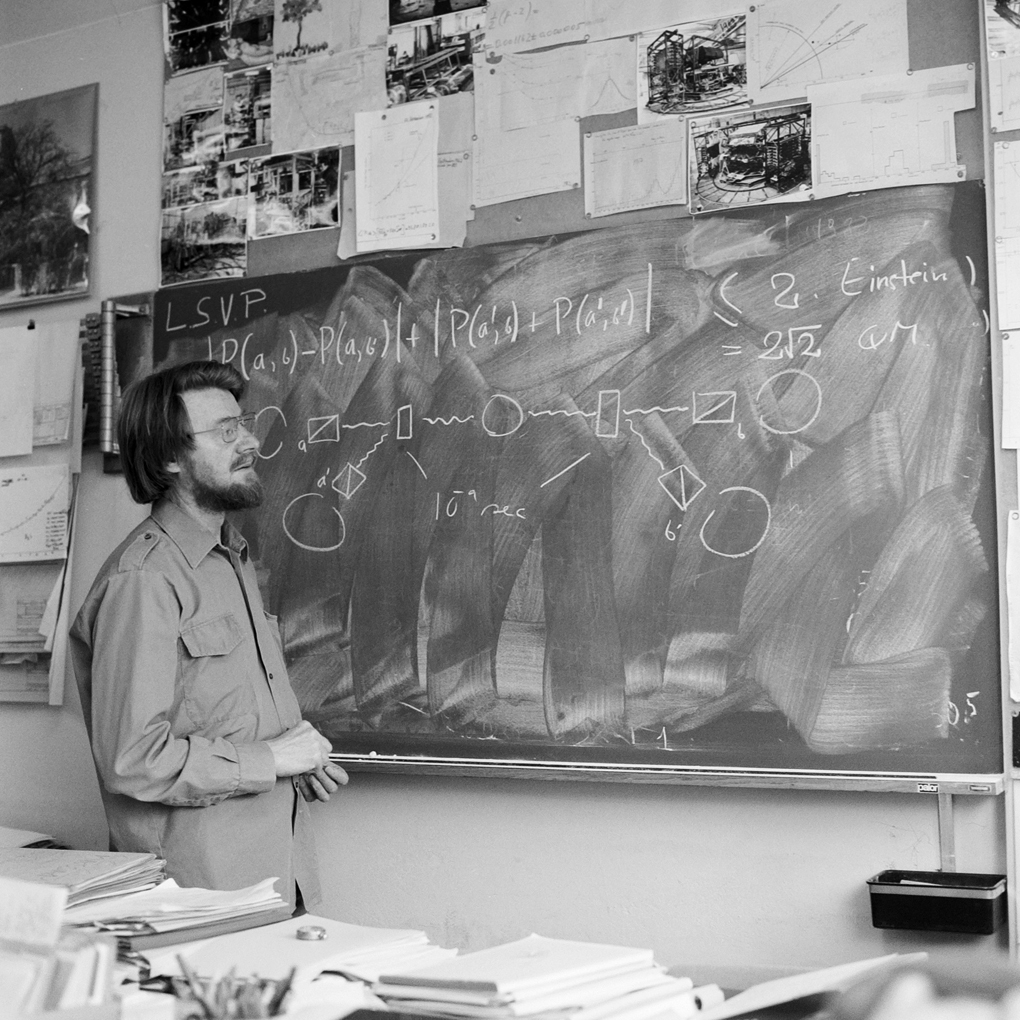}
\caption{\label{Fig1}John Bell at CERN. \copyright 1982 CERN.}
\end{figure}



\section{Introduction}


There is a photograph of John Bell taken in 1982 in front of a blackboard in his office at CERN. The famous Clauser-Horne-Shimony-Holt (CHSH) Bell inequality \cite{Bell64,CHSH69} is written in the blackboard. At the right hand side of the maximum bound for local hidden variable theories it is written ``Einstein.'' Below that, it is written the maximum for quantum theory (QT): ``$2 \sqrt{2}$.'' Already in 1969, CHSH noticed that this was the maximum for two-qubit systems \cite{CHSH69}. In 1980, Tsirelson proved that it is also the maximum in QT, no matter the dimensionality of the state space \cite{Tsirelson80}. It took a lot of time until somebody asked the obvious question: why? \cite{PR94}. It took a surprising amount of time until somebody came with a compelling answer \cite{PPKSWZ09}. However, it was soon proved that this answer cannot explain the maximum quantum values for some tripartite Bell inequalities \cite{YCATS12}. This leads us back to square one: Why the quantum maxima of all Bell inequalities? What is the fundamental reason that limits quantum probabilities?

In the summer of 1964, before submitting the Bell inequality paper, Bell submitted other paper which, for several reasons \cite{Jammer74,Jammer90}, was not published until 1966 \cite{Bell66}. There, Bell discusses the implications for the hidden variables problem of Gleason's theorem \cite{Gleason57}, which was directed to reducing the axiomatic basis of QT. The relevant corollary of Gleason's theorem is that, if the dimensionality $d$ of the state space is grater than two, then there exists a set of elementary tests such that values $t$(rue) or $f$(alse) cannot be assigned to them respecting that: (i) $t$ cannot be assigned to two mutually exclusive tests, and (ii) $t$ must be assigned to exactly one of $d$ mutually exclusive tests. Bell proved this corollary by constructing an explicit infinite set of elementary tests in $d=3$ for which such an assignment is impossible. A finite set was found by Kochen and Specker in 1962 \cite{Kochen12}, but not published until 1967 \cite{KS67}, making explicit a result anticipated by Specker in 1960 \cite{Specker60}. These sets prove the impossibility of reproducing QT with theories satisfying the assumption of non-contextuality of results, namely, the assumption that the ``measurement of an observable must yield the same value independently of what other measurements may be made simultaneously'' \cite{Bell66}. Bell considered that this assumption was not reasonable and finished his paper suggesting that it would be interesting to pursue some proof of impossibility of hidden variables replacing non-contextuality by some assumption of locality. One month later, Bell submitted the Bell inequality paper.

However, the problem of hidden variables in QT can be mathematically formulated in a way which goes beyond whether or not non-contextuality is reasonable. The problem of hidden variables in QT is simply whether or not it is possible to recover the quantum probabilities from a joint probability distribution over a single probability space. What is proven by Kochen-Specker and Bell's examples is that this is not possible in any scenario in which the dimensionality of the state space is three or higher, irrespective of whether or not locality can be invoked.


\section{The KCBS and the CHSH inequalities}


What if Bell would have derived a Bell-like inequality violated by quantum systems of dimension three? Such an inequality was introduced in 2008 and is called the Klyachko-Can-Binicio\u{g}lu-Shumovsky (KCBS) non-contextuality (NC) inequality \cite{KCBS08}. The KCBS inequality is the analog for quantum systems of dimension three of the CHSH inequality. The KCBS inequality is the simplest NC inequality violated by quantum systems of dimension three; the CHSH inequality is the simplest NC inequality violated by quantum systems of dimension four. Simplicity is here measured by the number of dichotomic observables used. The quantum violation of NC inequalities shows the impossibility to recover the quantum probabilities from a joint probability distribution over a single probability space.

The KCBS inequality says that, for any non-contextual hidden variable (NCHV) theory,
\begin{equation}
\label{kcbs}
 \kappa \stackrel{\mbox{\tiny{NCHV}}}{\leq} 3,
\end{equation}
with
\begin{equation}
\label{kappa}
 \kappa = \langle A_1 A_2 \rangle + \langle A_2 A_3 \rangle + \langle A_3 A_4 \rangle + \langle A_4 A_5 \rangle - \langle A_5 A_1 \rangle,
\end{equation}
where $A_i$ are observables with two possible results $-1$ and $+1$, and
\begin{equation}
 \label{mean}
 \langle A_j A_{j+1} \rangle = P(A_j+,A_{j+1}+)-P(A_j+,A_{j+1}-)-P(A_j-,A_{j+1}+)+P(A_j-,A_{j+1}-),
\end{equation}
where, e.g., $P(A_j+,A_{j+1}-)$ denotes the joint probability of obtaining $+1$ when measuring $A_j$ and $-1$ when measuring $A_{j+1}$. Probabilities in Eq.~(\ref{mean}) are assumed to be well defined no matter in which order $A_j$ and $A_{j+1}$ are measured. However, for $A_j$ and $A_{j+2}$ this may not be the case.

Similarly, the CHSH inequality says that, for any local hidden variables (LHV) theory,
\begin{equation}
\label{chsh}
 \beta \stackrel{\mbox{\tiny{LHV}}}{\leq} 2,
\end{equation}
with
\begin{equation}
\label{beta}
 \beta = \langle A_1 A_2 \rangle + \langle A_2 A_3 \rangle + \langle A_3 A_4 \rangle - \langle A_4 A_1 \rangle.
\end{equation}
The difference between $\kappa$ and $\beta$ is that, in $\beta$, $A_1$ and $A_3$ can be measured on a subsystem while $A_2$ and $A_4$ are measured on a distant subsystem. Therefore, in $\beta$ the choice of measurement on one subsystem and the result on the other subsystem can be space-like separated. This allows us to invoke locality to justify the assumption of non-contextuality.

In contrast with (\ref{kcbs}) and (\ref{chsh}), in QT,
\begin{equation}
 \kappa \stackrel{\mbox{\tiny{QT}}}{\leq} 4 \sqrt{5}-5 \approx 3.944
\end{equation}
and
\begin{equation}
 \beta \stackrel{\mbox{\tiny{QT}}}{\leq} 2 \sqrt{2} \approx 2.828.
\end{equation}
The big question is not just why QT violates the inequalities for hidden variable theories, but rather why QT violates them {\em exactly up to these limits}.


\section{Introducing the exclusivity principle}


Consider non-demolition measurements that are repeatable (i.e., that give the same result when repeated) and cause no disturbance on other measurements (i.e., when combined with these other measurements, all are repeatable). These measurements are called ``sharp'' \cite{Kleinmann14, CY14} and, in QT, are represented by projection operators. These are the measurements that von Neumann called ``quantum observables'' \cite{vonNeumann32}. Let us define an {\em event} as the state of the system after some sharp measurements (with certain results) on some initial state. Two events are equivalent when they correspond to indistinguishable states. Two events are exclusive when there is a measurement that distinguishes between them.

A theory satisfies the {\em exclusivity (E) principle} \cite{Cabello13} when any set of $n$ pairwise exclusive events is $n$-wise exclusive. Therefore, if we assume the E principle, Kolmogorov's axioms of probability lead us to the conclusion that the sum of the probabilities of any set of pairwise exclusive events cannot be higher than~$1$.

However, the E principle cannot be derived from Kolmogorov's axioms. To see it, consider the maximum value of the following sum of probabilities of events:
\begin{equation}
\label{spe}
 S = P(A_1+,A_2+)+P(A_2-,A_3-)+P(A_3+,A_1-),
\end{equation}
where the notation is the same used above. The three events $(A_1+,A_2+)$, $(A_2-,A_3-)$ and $(A_3+,A_1-)$ are pairwise exclusive. Therefore, the only restrictions from Kolmogorov's axioms are that the probabilities are non-negative and that
\begin{subequations}
\begin{align}
 P(A_1+,A_2+)+P(A_2-,A_3-) \le 1,\\
 P(A_2-,A_3-)+P(A_3+,A_1-) \le 1,\\
 P(A_3+,A_1-)+P(A_1+,A_2+) \le 1.
\end{align}
\end{subequations}
Therefore, for theories satisfying Kolmogorov's axioms the maximum is $S=3/2$, since each of the three probabilities in (\ref{spe}) can be $1/2$. However, for theories satisfying the E principle, the maximum is $S=1$, since the E principles forces that
\begin{equation}
 P(A_1+,A_2+)+P(A_2-,A_3-)+P(A_3+,A_1-) \stackrel{\mbox{\tiny{E}}}{\leq} 1.
\end{equation}

The E principle can be derived from a variety of axioms. For example, from the axiom that pairwise co-measurability implies joint co-measurability \cite{Specker60}, from the principle of fundamental sharpness of measurements \cite{CY14}, from axioms 1 and 2 in Ref.~\cite{BMU14}, and from the principle of lack of irreducible third order interference \cite{Henson14}.

The E principle imposes limits to the sum of probabilities of pairwise exclusive events. Therefore, in order to study the implications of the E principle for the limits of the KCBS and CHSH inequalities, it is convenient to rewrite both inequalities in terms of sums of probabilities of events. For that, it is useful to notice that the condition of normalization of probabilities allows us to write
\begin{subequations}
\begin{align}
 \langle A_j A_{j+1} \rangle &= 2 P(A_j+,A_{j+1}+)+2 P(A_j-,A_{j+1}-)-1,\\
 -\langle A_j A_{j+1} \rangle &= 2 P(A_j+,A_{j+1}-)+2 P(A_j-,A_{j+1}+)-1.
\end{align}
\end{subequations}
Therefore, we can write
\begin{subequations}
\begin{align}
 \kappa =& 2 S_{\rm KCBS} + 2 S'_{\rm KCBS} -5,\\
 \beta =& 2 S_{\rm CHSH} -4,
\end{align}
\end{subequations}
where
\begin{subequations}
\begin{align}
 S_{\rm KCBS}=&P(A_1+,A_2+)+P(A_2-,A_3-)+P(A_3+,A_4+)+P(A_4-,A_5-)+P(A_5+,A_1-),\\
 S_{\rm CHSH}=&P(A_1+,A_2+)+P(A_1-,A_2-)+P(A_2+,A_3+)+P(A_2-,A_3-)\nonumber \\
 &+P(A_3+,A_4+)+P(A_3-,A_4-)+P(A_4+,A_1-)+P(A_4-,A_1+)
\end{align}
\end{subequations}
and $S'_{\rm KCBS}$ is obtained from $S_{\rm KCBS}$ by changing the signs of all the results. Then, we can write the KCBS and CHSH inequalities and their quantum limits as follows:
\begin{subequations}
\begin{align}
 S_{\rm KCBS} &\stackrel{\mbox{\tiny{NCHV}}}{\leq} 2 \stackrel{\mbox{\tiny{QT}}}{\leq} \sqrt{5} \approx 2.236,\\
 S_{\rm CHSH} &\stackrel{\mbox{\tiny{LHV}}}{\leq} 3 \stackrel{\mbox{\tiny{QT}}}{\leq} 2+ \sqrt{2} \approx 3.414.
\end{align}
\end{subequations}


\section{The limit of the KCBS inequality}


Our first target is to explain why $S_{\rm KCBS}$ cannot go beyond $\sqrt{5}$ or, equivalently, why $\kappa$ cannot go beyond $4 \sqrt{5}-5$. For this purpose, consider two independent experiments both aiming the maximum of $S_{\rm KCBS}$. Suppose that one of the experiments is performed in Vienna on a certain physical system, while the other experiment is performed in Stockholm on a different physical system. Let us denote by $(A_j+,A_{j+1}+)$ an event of the experiment in Vienna, by $S_{\rm KCBS}^A$ the sum of the corresponding five probabilities, by $(B_k+,B_{k+1}+)$ an event of the experiment in Stockholm and by $S_{\rm KCBS}^B$ the sum of the corresponding five probabilities.

Since the experiments are independent, the probability of an event involving both experiments is the product of the probabilities of the corresponding (single-city) events. For example,
\begin{equation}
\label{indep}
 P(A_j+,A_{j+1}+,B_k+,B_{k+1}+)=P(A_j+,A_{j+1}+) P(B_k+,B_{k+1}+).
\end{equation}


\begin{figure}[t]
\centering
\includegraphics[scale=0.80]{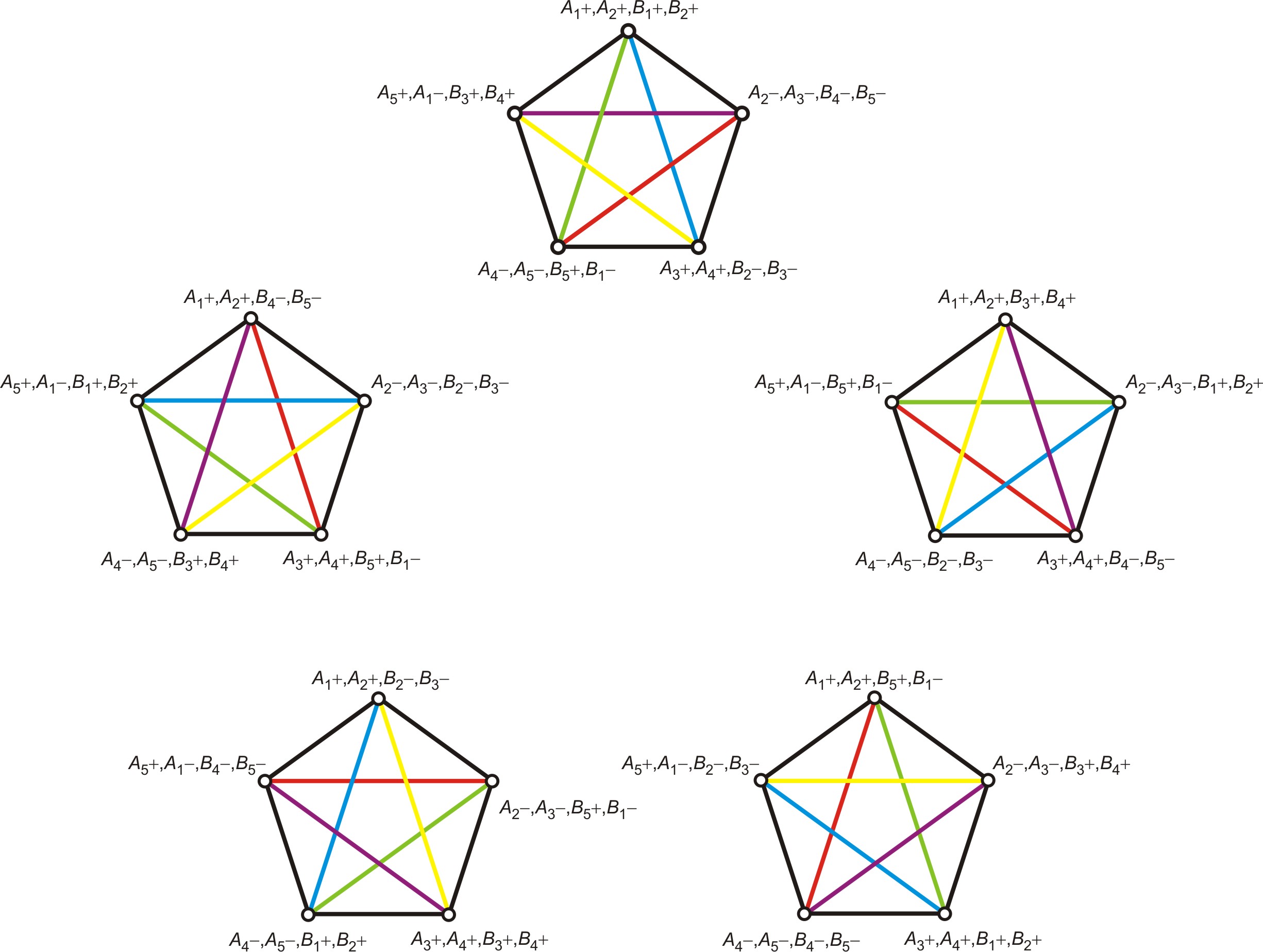}
\caption{\label{Fig2}Exclusivity graphs of the five sets of five pairwise exclusive events used in the proof of the limit of the KCBS inequality. Events are represented by nodes and exclusivity relations by edges. The black pentagons correspond to the exclusivity relations between the events $(A_j \gamma,A_{j+1} \delta)$. The coloured pentagrams correspond to the exclusivity relations between the events $(B_k \epsilon,B_{k+1} \phi)$. Any two graphs differ in a rotation of the pentagram.}
\end{figure}


Having two copies, we can identify larger sets of pairwise exclusive events. For example, the set with the following events: $(A_1+,A_2+,B_1+,B_2+)$, $(A_2-,A_3-,B_4-,B_5-)$, $(A_3+,A_4+,B_2-,B_3-)$, $(A_4-,A_5-,B_5+,B_1-)$ and $(A_5+,A_1-,B_3+,B_4+)$. The E principle and assumption (\ref{indep}) applied to this set imply that
\begin{subequations}
\begin{eqnarray}
\label{uno}
 &&P(A_1+,A_2+) P(B_1+,B_2+) + P(A_2-,A_3-) P(B_4-,B_5-) + P(A_3+,A_4+) P(B_2-,B_3-) \nonumber \\
 &&+ P(A_4-,A_5-) P(B_5+,B_1-) + P(A_5+,A_1-) P(B_3+,B_4+) \stackrel{\mbox{\tiny{E}}}{\leq} 1.
\end{eqnarray}
Similarly, by identifying sets of pairwise exclusive events, we can derive the following inequalities:
\begin{eqnarray}
 &&P(A_1+,A_2+) P(B_3+,B_4+) + P(A_2-,A_3-) P(B_1+,B_2+) + P(A_3+,A_4+) P(B_4-,B_5-) \nonumber \\
 &&+ P(A_4-,A_5-) P(B_2-,B_3-) + P(A_5+,A_1-) P(B_5+,B_1-) \stackrel{\mbox{\tiny{E}}}{\leq} 1,\\
 &&P(A_1+,A_2+) P(B_5+,B_1-) + P(A_2-,A_3-) P(B_3+,B_4+) + P(A_3+,A_4+) P(B_1+,B_2+) \nonumber \\
 &&+ P(A_4-,A_5-) P(B_4-,B_5-) + P(A_5+,A_1-) P(B_2-,B_3-) \stackrel{\mbox{\tiny{E}}}{\leq} 1,\\
 &&P(A_1+,A_2+) P(B_2-,B_3-) + P(A_2-,A_3-) P(B_5+,B_1-) + P(A_3+,A_4+) P(B_3+,B_4+) \nonumber \\
 &&+ P(A_4-,A_5-) P(B_1+,B_2+) + P(A_5+,A_1-) P(B_4-,B_5-) \stackrel{\mbox{\tiny{E}}}{\leq} 1,\\
 &&\label{cinco} P(A_1+,A_2+) P(B_4-,B_5-) + P(A_2-,A_3-) P(B_2-,B_3-) + P(A_3+,A_4+) P(B_5+,B_1-) \nonumber \\
 &&+ P(A_4-,A_5-) P(B_3+,B_4+) + P(A_5+,A_1-) P(B_1+,B_2+) \stackrel{\mbox{\tiny{E}}}{\leq} 1.
\end{eqnarray}
\end{subequations}
The geometry behind these sets is explained in Fig.~\ref{Fig2}.
If we sum all five inequalities (\ref{uno})--(\ref{cinco}), we obtain
\begin{equation}
 S_{\rm KCBS}^A S_{\rm KCBS}^B \stackrel{\mbox{\tiny{E}}}{\leq} 5.
\end{equation}
Assuming that the maximum is the same in both experiments, i.e., that $S_{\rm KCBS}^A = S_{\rm KCBS}^B$, we can conclude that, for any theory satisfying the E principle,
\begin{equation}
 S_{\rm KCBS} \stackrel{\mbox{\tiny{E}}}{\leq} \sqrt{5}.
\end{equation}
Exactly as in QT. This is an arguably clearer presentation of a result introduced in Ref.~\cite{Cabello13}.


\section{The limit of the CHSH inequality}


Our second target is to explain why $S_{\rm CHSH}$ cannot go beyond $2 + \sqrt{2}$ or, equivalently, why $\beta$ cannot go beyond $2 \sqrt{2}$. For this purpose, first notice that the state space on which $A_1$ and $A_3$ act is, at least, two-dimensional. Second, notice that the conditions of normalization of probabilities allows us to write,
\begin{eqnarray}
 4-S_{\rm CHSH}=&&P(A_1+,A_2-)+P(A_1-,A_2+)+P(A_2+,A_3-)+P(A_2-,A_3+)\nonumber \\
 &&+P(A_3+,A_4-)+P(A_3-,A_4+)+P(A_4+,A_1+)+P(A_4-,A_1-).
\end{eqnarray}

Now consider two independent experiments both testing $S_{\rm CHSH}$. Suppose that one of the experiments is performed in Vienna and the other experiment in Stockholm. As before, let us denote by $(A_j+,A_{j+1}+)$ one event in Vienna and by $(B_k+,B_{k+1}+)$ one event in Stockholm.

Now notice that $A_1$ and $B_1$ are co-measurable and that the state space on which $A_1$ and $B_1$ act is, at least, four-dimensional. Therefore, there must exist an observable $C_{11}$ co-measurable with $A_1$ and $B_1$ and such that the result for $C_{11}$ is $+1$ if the results of $A_1$ and $B_1$ are equal, and $-1$ if they are different. $C_{11}$ acts on a four-dimensional state space, but only distinguishes between two subspaces. Therefore, there must be an observable $C_{33}$ co-measurable with $C_{11}$ that distinguishes between other two subspaces and such that
\begin{equation}
\label{norma}
 P(C_{11}+,C_{33}+)+P(C_{11}+,C_{33}-)+P(C_{11}-,C_{33}+)+P(C_{11}-,C_{33}-)=1.
\end{equation}
Since we have made no assumption about $A_3$ and $B_3$ (other than each of them acts on a different at least two-dimensional subspace), we can relate $C_{33}$ to $A_3$ and $B_3$, the same way we related $C_{11}$ to $A_1$ and $B_1$. Similarly, we can start with $A_1$ and $B_3$ and define $C_{13}$ and then define a co-measurable $C_{31}$ related to $A_3$ and $B_1$.

These observables allow us to identify larger sets of pairwise exclusive events. For example, the set with the following events: $(A_1+,A_2+,B_1+,B_2+,C_{11}+)$, $(A_1+,A_2-,B_1+,B_2-,C_{11}+)$, $(A_3+,A_2+,B_3-,B_2-,C_{33}-)$, $(A_3+,A_2-,B_3-,B_2+,C_{33}-)$,
$(A_1-,A_2-,B_1-,B_2-,C_{11}+)$, $(A_1-,A_2+,B_1-,B_2+,C_{11}+)$, $(A_3-,A_2-,B_3+,B_2+,C_{33}-)$, $(A_3-,A_2+,B_3+,B_2-,C_{33}-)$ and $(C_{11}-,C_{33}+)$.
Since, by definition of $C_{11}$, $P(A_1+,A_2+,B_1+,B_2+,C_{11}+)=P(A_1+,A_2+,B_1+,B_2+)$, the E principle and assumption (\ref{indep}) applied to this set imply that
\begin{eqnarray}
\label{one}
 &&P(A_1+,A_2+) P(B_1+,B_2+) + P(A_1+,A_2-) P(B_1+,B_2-) + P(A_3+,A_2+) P(B_3-,B_2-) \nonumber \\
 &&+ P(A_3+,A_2-) P(B_3-,B_2+)+ P(A_1-,A_2-) P(B_1-,B_2-) + P(A_1-,A_2+) P(B_1-,B_2+) \nonumber \\
 &&+ P(A_3-,A_2-) P(B_3+,B_2+) + P(A_3-,A_2+) P(B_3+,B_2-) + P(C_{11}-,C_{33}+) \stackrel{\mbox{\tiny{E}}}{\leq} 1.
\end{eqnarray}


\begin{table*}[tb]
\begin{tabular}{ccccc}
\hline
\hline
$(A_1+,A_2+,B_1+,B_2+)$ &$(A_1+,A_2-,B_1+,B_2-)$ &$(A_3+,A_2+,B_3-,B_2-)$ &$(A_3+,A_2-,B_3-,B_2+)$ &$(C_{11}-,C_{33}+)$\\
$(A_1+,A_2+,B_1-,B_2-)$ &$(A_1+,A_2-,B_1-,B_2+)$ &$(A_3+,A_2+,B_3+,B_2+)$ &$(A_3+,A_2-,B_3+,B_2-)$ &$(C_{11}+,C_{33}-)$\\
$(A_1+,A_2+,B_1-,B_4-)$ &$(A_1+,A_2-,B_1-,B_4+)$ &$(A_3+,A_2+,B_3-,B_4+)$ &$(A_3+,A_2-,B_3-,B_4-)$ &$(C_{11}+,C_{33}+)$\\
$(A_1+,A_2+,B_1+,B_4+)$ &$(A_1+,A_2-,B_1+,B_4-)$ &$(A_3+,A_2+,B_3+,B_4-)$ &$(A_3+,A_2-,B_3+,B_4+)$ &$(C_{11}-,C_{33}-)$\\
$(A_1+,A_2+,B_3-,B_4+)$ &$(A_1+,A_2-,B_3-,B_4-)$ &$(A_3+,A_2+,B_1-,B_4-)$ &$(A_3+,A_2-,B_1-,B_4+)$ &$(C_{13}+,C_{31}+)$\\
$(A_1+,A_2+,B_3+,B_4-)$ &$(A_1+,A_2-,B_3+,B_4+)$ &$(A_3+,A_2+,B_1+,B_4+)$ &$(A_3+,A_2-,B_1+,B_4-)$ &$(C_{13}-,C_{31}-)$\\
$(A_1+,A_2+,B_3+,B_2+)$ &$(A_1+,A_2-,B_3+,B_2-)$ &$(A_3+,A_2+,B_1-,B_2-)$ &$(A_3+,A_2-,B_1-,B_2+)$ &$(C_{13}-,C_{31}+)$\\
$(A_1+,A_2+,B_3-,B_2-)$ &$(A_1+,A_2-,B_3-,B_2+)$ &$(A_3+,A_2+,B_1+,B_2+)$ &$(A_3+,A_2-,B_1+,B_2-)$ &$(C_{13}+,C_{31}-)$\\
\hline
\hline
\end{tabular}
\caption{Eight sets of nine pairwise exclusive events. Each row displays five pairwise exclusive events. For each row, there are other four events (not displayed) which are obtained by changing the results of the first four events. There are other eight sets (not displayed) which are obtained by exchanging $(A_j \gamma,A_k \delta,B_l \epsilon,B_m \phi)$ by $(A_l \epsilon,A_m \phi,B_j \gamma,B_k \delta)$.
\label{Table1}}
\end{table*}


As explained in Table~\ref{Table1}, there are 16 sets like this one. For each of them, there is a inequality like (\ref{one}). If we sum all of them we obtain,
\begin{equation}
 S_{\rm CHSH}^A S_{\rm CHSH}^B+(4-S_{\rm CHSH}^A) (4-S_{\rm CHSH}^B)+4 \stackrel{\mbox{\tiny{E}}}{\leq} 16.
\end{equation}
Assuming that the maximum is the same in both experiments, we can conclude that, for any theory satisfying the E principle,
\begin{equation}
 S_{\rm CHSH} \stackrel{\mbox{\tiny{E}}}{\leq} 2 + \sqrt{2}.
\end{equation}
Exactly as in QT. This result is an improved version of an argument introduced in Ref.~\cite{Cabello14}. A similar argument allows us to derive the quantum limits for $n$-partite Bell-like inequalities for non-local (but not genuinely $n$-partite non-local) hidden variable theories \cite{Cabello14b}.


\section{The unspeakable why}


We have shown that some characteristic limits of QT have a simple explanation. Indeed, we suspect that all the limits of quantum probabilities have the same explanation. If this would be the case, what should we learn about QT?

For some people, a fundamental message of QT is that the world is non-local, i.e., that the results of quantum observables correspond to some reality and change non-locally \cite{Gisin12}. However, from this perspective, it is puzzling that the no-signaling principle allows for higher than quantum violations of the CHSH inequality \cite{PR94}. Why then QT is not more non-local?

The reason why QT is exactly as non-local as it is, apparently, the same reason why QT is exactly as contextual as it is. However, in the contextuality case, there is no Alice and Bob and no communication.

Why nobody paid attention to the E principle before? Arguably, because the E principle is trivial in classical deterministic theories and not well defined in non-local realistic theories.

However, if one takes non-realism (of the results of observables represented in QT by self-adjoint operators) as a fundamental key of the world, then everything makes much more sense. The fundamental non-existence of results makes that not all conceivable combinations of observables allow for joint probability distributions (i.e., makes not all observables to be co-measurable). Indeed, it makes all conceivable relations of co-measurability/non-co-measurability (among sharp measurements) realizable and, as a consequence, makes all conceivable relationships of exclusivity/non-exclusivity (among events) realizable. There is where the E principle makes a profound contribution: The possible sets of probabilities of a given scenario (i.e., a certain structure of co-measurability/non-co-measurability) are restricted by the E principle applied to all conceivable embeddings of the scenario into a larger scenario. In this sense, the E principle acts in an holistic way. In particular, the limits of the probabilities of a given scenario follow from identifying the most (or one of the most) restrictive embedding(s) (details will be presented elsewhere \cite{Cabello15}). The resulting picture points out that non-realism is not ``a soft option'' \cite{Gisin12}, but rather a fundamental key of the world. QT is a probability theory about things that do not exist and are unpredictable at a fundamental level.

One may argue that the view I have drawn before is just one of the possible options and that the predictions of QT are also compatible with contextual and non-local realistic views of the world. I disagree. Common to all these other views is a certain degree of realism that ranges from hidden variables determining the results of all possible experiments to just taking the quantum state as real. It seems evident that any of these other views, when examined in detail, will lead to predictions that QT does not make. Hopefully, we will soon identify these predictions and test them. Time and experiments will tell.


\section*{Acknowledgments}


This work was supported by the project FIS2011-29400 (MINECO, Spain) with FEDER funds, the FQXi large grant project ``The Nature of Information in Sequential Quantum Measurements'' and the program Science without Borders (CAPES and CNPq, Brazil).




\begin{thebibliography}{00}


\bibitem{Bell64}
 J. S. Bell,
 Physics (Long Island City, N.Y.) \textbf{1}, 195 (1964).

\bibitem{CHSH69}
 J. F. Clauser, M. A. Horne, A. Shimony, and R. A. Holt,
 \href{http://dx.doi.org/10.1103/PhysRevLett.23.880}{Phys. Rev. Lett. \textbf{23}, 880 (1969).}

\bibitem{Tsirelson80}
 B. S. Cirel'son [Tsirelson],
 \href{http://dx.doi.org/10.1007/BF00417500}{Lett. Math. Phys. \textbf{4}, 93 (1980).}

\bibitem{PR94}
 S. Popescu and D. Rohrlich,
 \href{http://link.springer.com/article/10.1007%2FBF02058098}{Found. Phys. \textbf{24}, 379 (1994).}

\bibitem{PPKSWZ09}
 M. Paw{\l}owski, T. Paterek, D. Kaszlikowski, V. Scarani, A. Winter, and M. \.{Z}ukowski,
 \href{http://www.nature.com/nature/journal/v461/n7267/full/nature08400.html}{Nature \textbf{461}, 1101 (2009).}

\bibitem{YCATS12}
 T. H. Yang, D. Cavalcanti, M. L. Almeida, C. Teo, and V. Scarani,
 \href{http://iopscience.iop.org/1367-2630/14/1/013061}{New J. Phys. \textbf{14}, 013061 (2012).}

\bibitem{Jammer74}
 M. Jammer,
 {\em The Philosophy of Quantum Mechanics} (Wiley, New York, 1974).

\bibitem{Jammer90}
 M. Jammer,
 \href{http://link.springer.com/article/10.1007%2FBF01889462}{Found. Phys. \textbf{20}, 1139 (1990).}

\bibitem{Bell66}
 J. S. Bell,
 \href{http://dx.doi.org/10.1103/RevModPhys.38.447}{Rev. Mod. Phys. \textbf{38}, 447 (1966).}

\bibitem{Gleason57}
 A. M. Gleason,
 J. Math. Mech. \textbf{6}, 885 (1957).

\bibitem{Kochen12}
 S. Kochen,
 in E. Engeler, N. Hungerb\"uhler, and J. A. Makowsky (eds.),
 \href{http://www.math.ethz.ch/~engeler/Specker_Nachruf.pdf}{Elem. Math. \textbf{67}, 1 (2012).}

\bibitem{KS67}
 S. Kochen and E. P. Specker,
 J. Math. Mech. \textbf{17}, 59 (1967).

\bibitem{Specker60}
 E. P. Specker,
 \href{http://onlinelibrary.wiley.com/doi/10.1111/j.1746-8361.1960.tb00422.x/abstract}{Dialectica \textbf{14}, 239 (1960).}
 English translation: \href{http://arxiv.org/abs/1103.4537}{\eprint{arXiv:1103.4537}.}

\bibitem{KCBS08}
 A. A. Klyachko, M. A. Can, S. Binicio\u{g}lu, and A. S. Shumovsky,
 \href{http://dx.doi.org/10.1103/PhysRevLett.101.020403}{Phys. Rev. Lett. \textbf{101}, 020403 (2008).}

\bibitem{Kleinmann14}
 M. Kleinmann,
 \href{http://dx.doi.org/10.1088/1751-8113/47/45/455304}{J. Phys. A: Math. Theor. \textbf{47}, 455304 (2014).}

\bibitem{CY14}
 G. Chiribella and X. Yuan,
 \href{http://arxiv.org/abs/1404.3348}{\eprint{arXiv:1404.3348}.}

\bibitem{vonNeumann32}
 J. von Neumann,
 {\em Mathematische Grundlagen der Quantenmechanik}
 (Springer-Verlag, Berlin, 1932).
 English version: {\em Mathematical Foundations of Quantum Mechanics}
 (Princeton University Press, Princeton, New Jersey, 1955).

\bibitem{Cabello13}
 A. Cabello,
 \href{http://dx.doi.org/10.1103/PhysRevLett.110.060402}{Phys. Rev. Lett. \textbf{110}, 060402 (2013).}
 
\bibitem{BMU14}
 H. Barnum, M. P. M\"uller, and C. Ududec,
 \href{http://dx.doi.org/10.1088/1367-2630/16/12/123029}{New J. Phys \textbf{16}, 123029 (2014).} 
 
\bibitem{Henson14}
 J. Henson,
 \href{http://dx.doi.org/10.1103/PhysRevLett.114.220403}{Phys. Rev. Lett. \textbf{114}, 220403 (2015).} 

\bibitem{Cabello14}
 A. Cabello,
 \href{http://dx.doi.org/10.1103/PhysRevA.90.062125}{Phys. Rev. A \textbf{90}, 062125 (2014).}
 
\bibitem{Cabello14b}
 A. Cabello,
 \href{http://dx.doi.org/10.1103/PhysRevLett.114.220402}{Phys. Rev. Lett. \textbf{114}, 220402 (2015).} 

\bibitem{Cabello15}
 A. Cabello
 (in preparation).

\bibitem{Gisin12}
 N. Gisin,
 \href{http://link.springer.com/article/10.1007%2Fs10701-010-9508-1#}{Found. Phys. \textbf{42}, 80 (2012).}


\end{thebibliography}
\end{document}